\documentstyle[10pt,epsfig,twocolumn]{article}
\newcommand{\sfbf}{\upshape\bfseries\sffamily}
\newcommand{\secf}{\large\sfbf}\newcommand{\subf}{\normalsize\sfbf}
\newcommand{\titf}{\LARGE\sfbf}\newfont{\calf}{cmfi10}
 
\newcommand{\head}{ {\calf\today} 
 \hfill \raisebox{-13mm}{\psfig{file=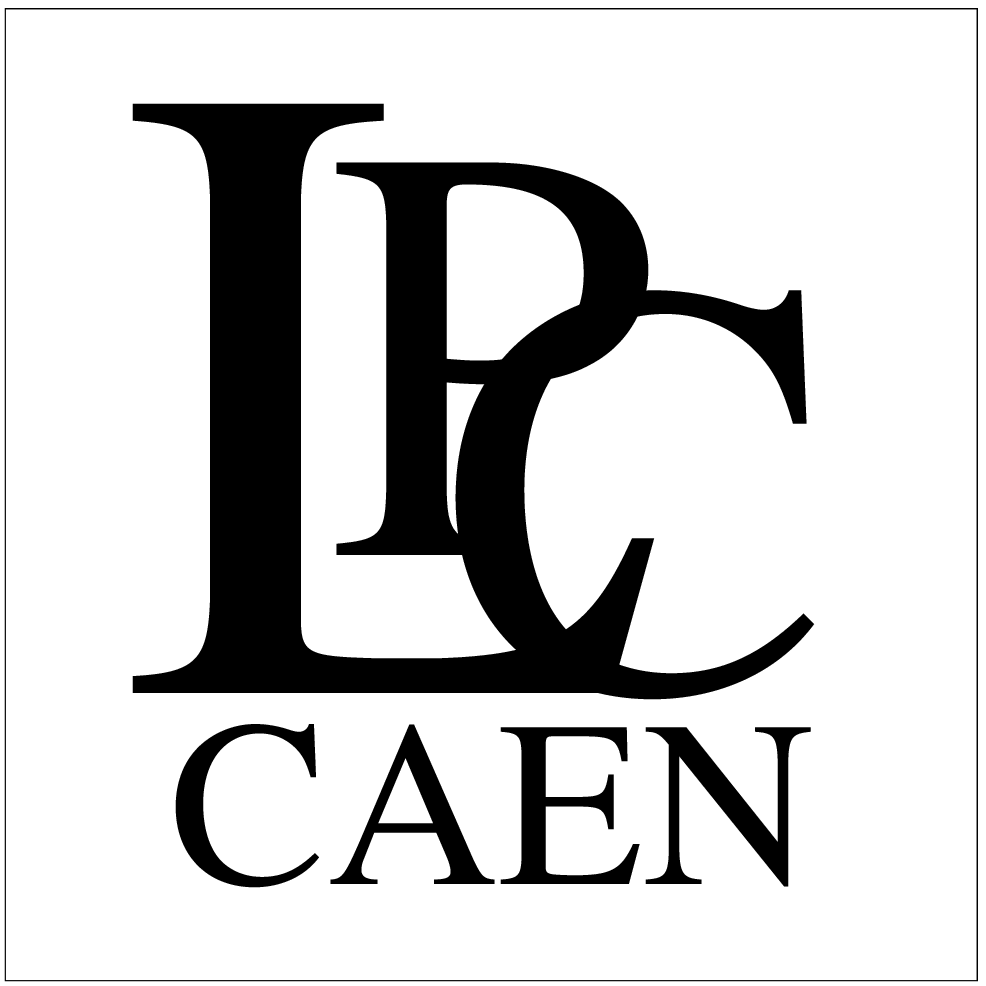,height=15mm}}}
\renewcommand{\title}[1]{\smallskip{\titf#1}\bigskip}
\def\ifdef#1#2#3{\expandafter\ifx\csname#1\endcsname\relax#3\else#2\fi}
\newcounter{Nad}
\renewcommand{\author}[2]{\ifdef{#1}{}{\expandafter\def\csname#1\endcsname{}
 \newcounter{#1}\stepcounter{Nad}\setcounter{#1}{\theNad}}{#2}$^{\arabic{#1}}$}
\newcommand{\address}[2]{{\small\it$^{\arabic{#1}}$~#2}\\}
\newcounter{Nth}\renewcommand{\thefootnote}{\alph{footnote}}
\renewcommand{\thanks}[1]
 {\stepcounter{Nth}\expandafter\gdef\csname thanks\roman{Nth}\endcsname
 {\stepcounter{Nth}\footnotetext[\value{NTH}]{~#1}}$^{,}$\footnotemark}
\newcounter{NTH}
\newcommand{\Thanks}{\stepcounter{NTH}\ifdef{thanks\roman{NTH}}
  {\csname thanks\roman{NTH}\endcsname\Thanks}
  {\setcounter{footnote}{0}\renewcommand{\thefootnote}{\fnsymbol{footnote}}} }
\renewenvironment{abstract}{\begin{minipage}{15cm}\small\hspace*{2mm}}
                           {\end{minipage}\bigskip\bigskip}
\newcommand{\ack}[1]{\bigskip\noindent{\small#1}\bigskip}
\let\Ocap\caption \renewcommand{\caption}[1]{\Ocap{\small#1}}
\let\Osec\section \renewcommand{\section}[1]{\Osec{\secf#1}}
\let\Osub\subsection \renewcommand{\subsection}[1]{\Osub{\subf#1}}

\newenvironment{references}{\let\section\Osec \begin{thebibliography}{99}}
               {\end{thebibliography}} 
 
\newcommand{\etal}{{\em et al.}}
\newcommand{\Epn}{E_{\rm{p}}/E_{\rm{n}}}
 
\begin{document}
 
\twocolumn[\head\begin{center}
 
\title{The detection of neutron clusters}
 
\author{LPC}	{F.M.~Marqu\'es}\thanks{Email: {\tt Marques@caelav.in2p3.fr}},
\author{LPC}	{M.~Labiche}\thanks{Present address:
		 University of Paisley, Scotland.},
\author{LPC}	{N.A.~Orr},
\author{LPC}	{J.C.~Ang\'elique},
\author{Chalm}	{L.~Axelsson},
\author{ULB}	{B.~Benoit},
\author{Aarhus}	{U.C.~Bergmann},
\author{CSIC}	{M.J.G.~Borge},
\author{Surrey}	{W.N.~Catford},
\author{Oxford}	{S.P.G.~Chappell},
\author{Bham}	{N.M.~Clarke},
\author{Ires}	{G.~Costa},
\author{Surrey}	{N.~Curtis}\thanks{Present address:
		 University of Birmingham, UK.},
\author{ULB}	{A.~D'Arrigo},
\author{ULB}	{E.~de~G\'{o}es~Brennand},
\author{GANIL}	{F.~de~Oliveira~Santos},
\author{Ires}	{O.~Dorvaux},
\author{Mess}	{G.~Fazio},
\author{LPC}	{M.~Freer$^{\arabic{Bham},}$},
\author{Bham}	{B.R.~Fulton}\thanks{Present address: University of York, UK.},
\author{Mess}	{G.~Giardina},
\author{IPNO}	{S.~Gr\'evy}\thanks{Present address: LPC, Caen, France.},
\author{IPNO}	{D.~Guillemaud-Mueller},
\author{ULB}	{F.~Hanappe},
\author{Ires}	{B.~Heusch},
\author{Chalm}	{B.~Jonson},
\author{LPC}	{C.~Le~Brun}\thanks{Present address: ISN, Grenoble, France.},
\author{IPNO}	{S.~Leenhardt},
\author{GANIL}	{M.~Lewitowicz},
\author{GANIL}	{M.J.~L\'opez}\thanks{Present address:
		 CEA-DAM, Bruy\`eres-le-Ch\^atel, France.},
\author{Chalm}	{K.~Markenroth},
\author{IPNO}	{A.C.~Mueller},
\author{Chalm}	{T.~Nilsson}\thanks{Present address:
		 ISOLDE, CERN, Switzerland.},
\author{LPC}	{A.~Ninane}\thanks{On leave from:
		 UCL-Louvain-la-Neuve, Belgium.},
\author{Chalm}	{G.~Nyman},
\author{CSIC}	{I.~Piqueras},
\author{Aarhus}	{K.~Riisager},
\author{GANIL}	{M.G.~Saint~Laurent},
\author{GANIL}	{F.~Sarazin}\thanks{Present address:
		 University of Edinburgh, Scotland.},
\author{Bham}	{S.M.~Singer},
\author{IPNO}	{O.~Sorlin},
\author{Ires}	{L.~Stuttg\'e}
 
\bigskip
 
\address{LPC}{Laboratoire de Physique Corpusculaire,
 IN2P3-CNRS, ISMRa et Universit\'e de Caen, F-14050 Caen cedex, France}
\address{Chalm}{Experimentell Fysik,
 Chalmers Tekniska H\"ogskola, S-412 96 G\"{o}teborg, Sweden}
\address{ULB}{Universit\'e Libre de Bruxelles,
 CP 226, B-1050 Bruxelles, Belgium}
\address{Aarhus}{Institut for Fysik og Astronomi,
 Aarhus Universitet, DK-8000 Aarhus C, Denmark}
\address{CSIC}{Instituto de Estructura de la Materia,
 CSIC, E-28006 Madrid, Spain}
\address{Surrey}{Department of Physics,
 University of Surrey, Guildford, Surrey, GU2~7XH, U.K.}
\address{Oxford}{Department of Nuclear Physics,
 University of Oxford, Keble Road, Oxford OX1 3RH, U.K.}
\address{Bham}{School of Physics and Astronomy,
 University of Birmingham, Birmingham B15 2TT, U.K.}
\address{Ires}{Institut de Recherche Subatomique, IN2P3-CNRS,
 Universit\'e Louis Pasteur, BP 28, F-67037 Strasbourg cedex, France}
\address{GANIL}{GANIL,
 CEA/DSM-CNRS/IN2P3, BP 55027, F-14076 Caen cedex, France}
\address{Mess}{Dipartimento di Fisica,
 Universit\`a di Messina, Salita Sperone 31, I-98166 Messina, Italy}
\address{IPNO}{Institut de Physique Nucl\'eaire,
 IN2P3-CNRS, F-91406 Orsay cedex, France}
 
\bigskip
 
\begin{abstract}
A new approach to the production and detection of bound neutron clusters is
presented. The technique is based on the breakup of beams of very neutron-rich
nuclei and the subsequent detection of the recoiling proton in a liquid
scintillator. The method has been tested in the breakup of $^{11}$Li, $^{14}$Be
and $^{15}$B beams by a C target. Some 6 events were observed that exhibit the
characteristics of a multineutron cluster liberated in the breakup of
$^{14}$Be, most probably in the channel $^{10}$Be+$^4$n. The various
backgrounds that may mimic such a signal are discussed in detail.
\end{abstract}
 
\end{center}]

\Thanks
 
\section{Introduction}
 
The very light nuclei have long played a fundamental role in testing nuclear
models and the underlying nucleon-nucleon interaction. Whilst much effort has
been expended in attempting to model stable systems, a number of ambiguities
remain. For example, the ground states of $^3$H and $^{3,4}$He do not appear to
be particularly sensitive to the form of the interaction \cite{Cie99}. In this
context the study of systems exhibiting very asymmetric $N/Z$ ratios may
provide new perspectives on the N-N interaction and few-body forces. In the
case of the light, two-neutron halo nuclei such as $^6$He, insight is already
being gained into the effects of the three-body force \cite{Zhu93}. Very
recently evidence has been presented that the ground state of $^5$H exists as a
relatively narrow, low-lying resonance \cite{Kor01}. In the case of the
lightest $N=4$ isotone, $^4$n, nothing is known \cite{Til92,Ogl89}. The
discovery of such neutral systems as bound states would have far reaching
implications for many facets of nuclear physics. In the present paper, the
production and detection of free neutron clusters is discussed.
 
The question whether neutral nuclei may exist has a long and checkered history
that may be traced back to the early 1960s \cite{Ogl89}. Forty years later the
only clear evidence in this respect is that the dineutron is particle unstable.
Although $^3$n is the simplest multineutron candidate, the effects of pairing
observed on the neutron drip-line suggest that $^{4,6,8}$n could exhibit bound
states \cite{Cie97}. Concerning the tetraneutron, an upper limit on the binding
energy of 3.1~MeV is provided by the particle stability of $^8$He, which does
not decay into $\alpha$+$^4$n. Furthermore, if $^4$n was bound by more than
1~MeV, $\alpha$+$^4$n would be the first particle threshold in $^8$He. As the
breakup of $^8$He is dominated by the $^6$He channel \cite{War00}, the
tetraneutron, if bound, should be so by less than 1~MeV.
 
The majority of the calculations performed to date suggest that multineutron
systems are unbound \cite{Til92}. Interestingly, it was also found that subtle
changes in the N-N potentials that do not affect the phase shift analyses may
generate bound neutron clusters \cite{Ogl89}. In addition to the complexity of
such {\em ab initio\/} calculations, which include the uncertainties in
many-body forces, the n-n interaction is the most poorly known N-N interaction,
as demonstrated by the controversy regarding the determination of the
$a_{\rm{nn}}$ scattering length \cite{Huh00}. The lack of predictive power of
calculations of few-body systems at the 1~MeV level (see for example
Ref.~\cite{Bev86}) therefore does not exclude the possible existence of a very
weakly bound $^4$n.
 
Experimentally, no limit has yet been placed on the binding energy of $^4$n (or
any other $^A$n). Rather only limits on the production cross-sections could be
estimated from two-step \cite{Sch63,Cie65,Det77,Boe80} or direct reactions
\cite{Ung84,Gor89,Gra99,Ohl68,Cer74,Bel88,Boh95}. In this paper, a new
technique to produce and detect neutron clusters is presented. The method is
based on the breakup of an energetic beam of a very neutron-rich nucleus and
the subsequent detection of the $^A$n cluster in a liquid scintillator. This
work represents the continuation of an experimental programme investigating the
structure and, in particular, the correlations within two-neutron halo systems
\cite{FMM96,FMM00,Lab01,FMM01}.
 
The paper is organised as follows. Section~2 describes the experimental
technique and analysis procedures. The results, including the observation of
some 6 events exhibiting characteristics consistent with a multineutron cluster
liberated in the breakup of $^{14}$Be, are presented in section~3. A detailed
discussion of the results is given in section~4. Special attention is paid to
the various backgrounds (most notably pile-up) that may mimic the signal
arising from a multineutron cluster. Finally a summary and outlook on future
work is briefly presented in section~5.
 
\section{Experimental technique}
 
\subsection{Previous experiments}
 
There have been essentially two categories of experiments that have searched
for $^A$n systems. The first consists of the production of $^A$n in reactions
such as neutron-induced fission of U \cite{Sch63,Cie65} or proton and light-ion
fragmentation of a heavy target \cite{Det77,Boe80}. Any recoiling $^A$n are
then, in principle, signaled by the radiochemical separation of decay products
from ($^A$n,$x$n) reactions in a secondary target. An extremely pure target and
a detailed analysis of all possible backgrounds are thus needed. As such only
upper limits for the $^A$n production cross-section, assuming cross-sections
for the ($^A$n,$x$n) reactions, could be determined. The only positive claim
\cite{Det77} was later explained as arising from an underestimation of the
production of very energetic tritons \cite{Boe80}.
 
The second class of experiments involves direct reactions of the type
$a(b,c)^A$n, where discrete values of the energy of the ejectile, $c$,
correspond to states in the $^A$n system. This technique can thus also reveal
unbound states. These searches have relied on the very low cross-section
(typically $\sim1$~nb) double-pion charge exchange (D$\pi$CX)
\cite{Ung84,Gor89,Gra99} and heavy-ion multinucleon-transfer reactions
\cite{Ohl68,Cer74,Bel88,Boh95}. Again no conclusive evidence for a bound $^A$n
or resonant states has been found in these studies. There are many problems
inherent in this technique: the precise knowledge of the many-body ($A$+1)
phase space; the background from target impurities; and the bias introduced by
the fact that both the $^A$n {\em and\/} the ejectile have to be formed in the
reaction, lowering further the production cross-section.
 
\subsection{Principle}
 
Clustering appears in many light nuclei close to particle emission thresholds
\cite{Fre99}. Examples include $\alpha$+t clustering in $^7$Li (threshold at
2.5~MeV), $\alpha$+$\alpha$+n in $^9$Be ($S_{\rm{n}}=1.6$~MeV), and
$\alpha$+4n+$\alpha$ in $^{12}$Be (threshold at 12.1~MeV). In light
neutron-rich nuclei, components of the wave function in which the neutrons
present a cluster-like configuration may be expected to appear \cite{Kan99}.
Owing to pairing and the confining effects of any underlying
$\alpha$-clustering, the most promising candidates may be the drip-line
isotopes of Helium and Beryllium, $^8$He ($S_{\rm4n}=3.1$~MeV) and $^{14}$Be
($S_{\rm4n}=5$~MeV).
 
We have investigated existing data for the breakup of a 35~MeV/N $^{14}$Be beam
by a C target \cite{FMM00,Lab01,Lab99}. Other components present in the beam
which will be exploited here were $^{11}$Li at 30~MeV/N and $^{15}$B at
48~MeV/N. In such reactions relatively high cross-sections (typically
$\sim100$~mb) are encountered. Consequently, even only a small component of the
wave function corresponding to a multineutron cluster could result in a
measurable yield with a moderate secondary beam intensity. Furthermore, the
backgrounds encountered in D$\pi$CX and heavy-ion transfer reactions are
obviated in direct breakup. The main difficulty in the approach lies in the
direct detection of the $^A$n cluster.
 
The details of the experimental setup have been described elsewhere
\cite{FMM00,Lab01,Lab99,fmm00}. Therefore only the salient features are
recalled. Following the breakup, the charged fragments were detected using
position-sensitive Si-CsI telescopes and the neutrons using 90 modules of the
DEMON array, located at distances of 3.5 to 6.5~m downstream of the target
\cite{Lab01}. The energy of the neutrons ($E_{\rm{n}}$) was derived from the
time-of-flight, the start being furnished by a thin Si detector placed just
forward of the target. An average resolution of 1.5~ns was obtained. For
particles with $A\geq2$, $E_{\rm{n}}$ corresponds to the energy per nucleon.
Standard pulse-shape discrimination methods based on the light output from the
liquid scintillator were employed to identify the neutrons from the $\gamma$
and cosmic-ray backgrounds \cite{Til95}. Except where noted, only events in
which precisely one DEMON module fired in coincidence with a charged fragment
were considered in the analysis, in order to avoid any possible contributions
from cross-talk \cite{fmm00}.
 
\begin{figure}
\begin{center}
 \mbox{\psfig{file=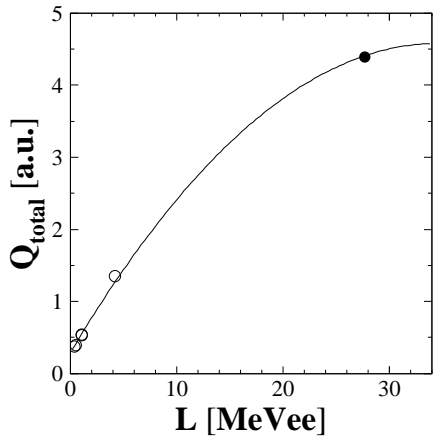,width=8.8cm}}
 \caption{Light output calibration for a DEMON module. The open symbols
correspond to the Compton edge of $\gamma$-rays from $^{22}$Na (511 and
1275~keV), $^{137}$Cs (662~keV), $^{60}$Co (1333~keV) and $^{241}$Am$^9$Be
(4.44~MeV) sources, and the closed symbol to cosmic-ray muons
\protect\cite{Lab99,Til95}. The solid line is a parabolic fit.} \label{f:QvL}
\end{center}
\end{figure}
 
\begin{figure}
\begin{center}
 \mbox{\psfig{file=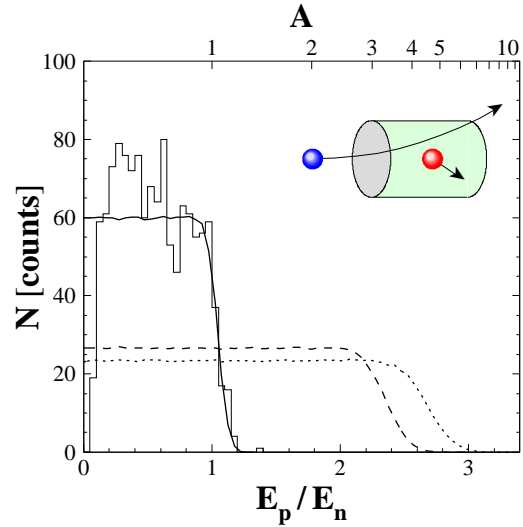,width=8.8cm}}
 \caption{Distribution of the ratio of proton energy, $E_{\rm{p}}$
(MeV), to the energy derived from the flight time, $E_{\rm{n}}$ (MeV/N), for
data from the reaction C($^{14}$Be,$^{12}$Be+n) (histogram) and for simulations
of elastic scattering of $^{1,3,4}$n (solid, dashed and dotted lines,
respectively) on protons. The experimental resolution has been included in the
simulations.} \label{f:ela}
\end{center}
\end{figure}

The predominant mechanism for the detection of neutrons in a liquid
scintillator such as that used in DEMON is n-p scattering \cite{Wan97}, in
which the proton recoils with an energy ($E_{\rm{p}}$) up to that of the
incident neutron. In general, the neutron does not lose all its energy in the
interaction and may escape from the detector \cite{fmm00}. The energy of the
recoiling proton can be determined from careful source and cosmic-ray
calibrations of the charge deposited in the module \cite{Lab99,Til95,Wan97}
(Fig.~\ref{f:QvL}). This may then be compared to the energy per nucleon of the
incident particle derived from the time-of-flight ($E_{\rm{n}}$). For a single
neutron and an ideal detector $\Epn\leq1$. For a real detector the finite
resolutions can give a higher limit, and for DEMON this is $\sim1.4$. In the
case of a multineutron cluster, $E_{\rm{p}}$ can exceed the incident energy per
nucleon and $\Epn$ may take on a range of values extending beyond 1.4, as shown
in Fig.~\ref{f:ela} ---the scale on the upper axis indicates the maximum value
as a function of the multineutron mass number.
 
Reactions on Carbon in the liquid scintillator, as for example n-C scattering
or C(n,3$\alpha$), do not present any difficulties to the present technique as
the associated light outputs translate to very low $\Epn$ \cite{Wan97}. In
addition, except for n-C scattering, the cross-sections for reactions on C are
negligible compared to scattering on Hydrogen in the energy range considered
here.
 
\subsection{Calibrations and energy range}
 
As is evident in the source and cosmic-ray calibrations in Fig.~\ref{f:QvL},
the DEMON modules exhibit saturation effects at very high light output, as
observed in earlier work \cite{Til95}. It should be noted that the initial goal
of the experiment run to acquire the data analysed here was not the search for
multineutron clusters, and as such the analysis of very high light outputs was
not foreseen\footnote{~The possibility of operating the DEMON photomultipliers
at lower voltages is being explored for future dedicated experiments.}. The
total charge versus light output is well described up to $\sim25$~MeVee using a
parabolic adjustment (Fig.~\ref{f:QvL}), as found in previous tests which
included measurements with a 15.5~MeV $\gamma$-ray \cite{Til95}. A deposited
charge of $\sim25$~MeVee corresponds to a proton recoil energy of
$E_{\rm{p}}\sim~32$~MeV. In order to avoid the effects of saturation,
particularly in the region $\Epn>1$, an upper limit of $E_{\rm{n}}=18$~MeV/N
has been imposed.
 
\begin{figure}[t]
\begin{center}
 \mbox{\psfig{file=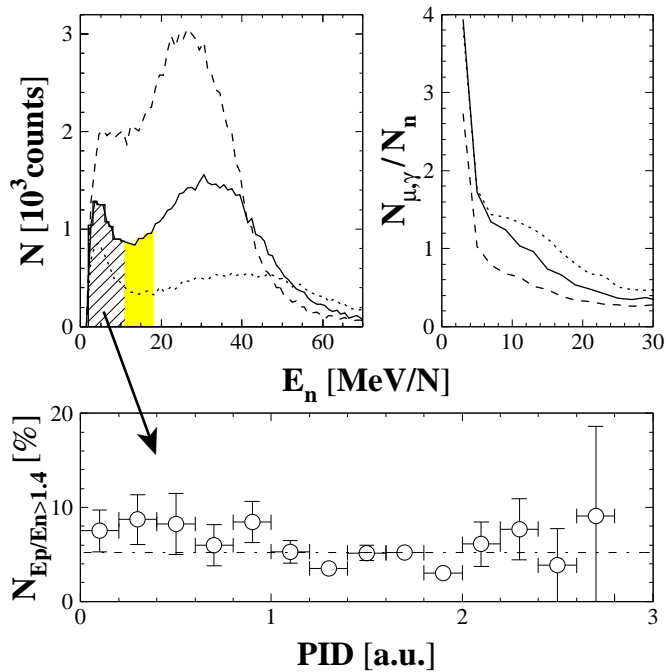,width=8.8cm}}
 \caption{Data from the reactions C($^AZ$,X+n) with $^{14}$Be, $^{11}$Li
and $^{15}$B beams (solid, dashed and dotted lines, respectively). Left-top:
neutron energy distribution; the shaded area corresponds to the energy range
used in the present analysis; the percentage of events in the hatched area with
$\Epn>1.4$ is shown in the lower panel as a function of the particle
identification parameter defined in Eq.~(\protect\ref{e:PID}). Right-top:
evolution with energy of the ratio of $\gamma$ and cosmic-ray events to
neutrons detected in DEMON.} \label{f:Ens}
\end{center}
\end{figure}
 
At low energies the proton recoil is free of saturation effects. However,
background events arising from $\gamma$ and cosmic rays represent a potential
contaminant of the $\Epn$ distribution. These events are randomly distributed
in time, and thus the relative rate increases at low energy (Fig.~\ref{f:Ens})
since $E{\propto}t^{-2}$. As the energy loss in a module is completely
uncorrelated with the inferred time-of-flight from the reaction at the target,
$\Epn$ is not confined below 1.4. Even if the rejection rate using pulse-shape
analysis is close to 100~\%, any events that remain could mimic a $^A$n signal.
In order to reject these events, we have first verified that their relative
rate at low energy is, as expected for a background, independent of the
reaction channel (lower panel in Fig.~\ref{f:Ens}). A lower limit on
$E_{\rm{n}}$ was then imposed and raised up to the value (11~MeV/N) for which
no events remain in the ($^{14}$Be,$^{12}$Be+n) channel, where no $^A$n should
be produced.
 
The energy gate, $E_{\rm{n}}=11$--18~MeV/N, is shown as the shaded area in
Fig.~\ref{f:Ens}. As may also be seen in this figure, the neutron energy
distributions exhibit two components: the neutrons from the breakup of the
projectile (distribution centred close to the beam velocity, $\sim30$~MeV/N)
and low energy neutrons evaporated by the excited target-like residue. In the
case of $^{15}$B, the neutrons arising from breakup are shifted to higher
energies due to the higher energy of the beam, and therefore the ratio of
background to neutrons is still relatively high at 11~MeV/N (dotted lines in
Fig.~\ref{f:Ens}). A limit of $E_{\rm{n}}=15$--18~MeV/N was thus imposed for
the $^{15}$B data.
 
\begin{figure}
\begin{center}
 \mbox{\psfig{file=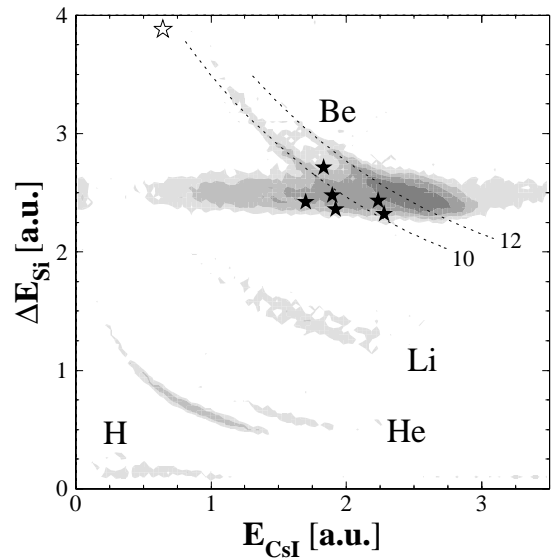,width=8.8cm}}
 \caption{Scatter plot of the energies deposited in the Si-CsI telescope
for the reaction C($^{14}$Be,X+n). Symbols correspond to the 7 events in
Fig.~\protect\ref{f:IvE} with $\Epn>1.4$. The horizontal band is discussed in
the text.} \label{f:loc}
\end{center}
\end{figure}
 
\section{Results}
 
The detection of neutrons produced in the reaction C($^{14}$Be,$^{12}$Be+n) is
displayed in Fig.~\ref{f:ela} (histogram); a channel in which $^A$n clusters
should be absent. We observe that the flat distribution predicted for n-p
scattering describes the data well, except for a small fraction of events at
low $\Epn$. As noted earlier, these correspond to reactions on $^{12}$C which
always generate smaller light outputs \cite{Wan97}.
 
\begin{figure}
\begin{center}
 \mbox{\psfig{file=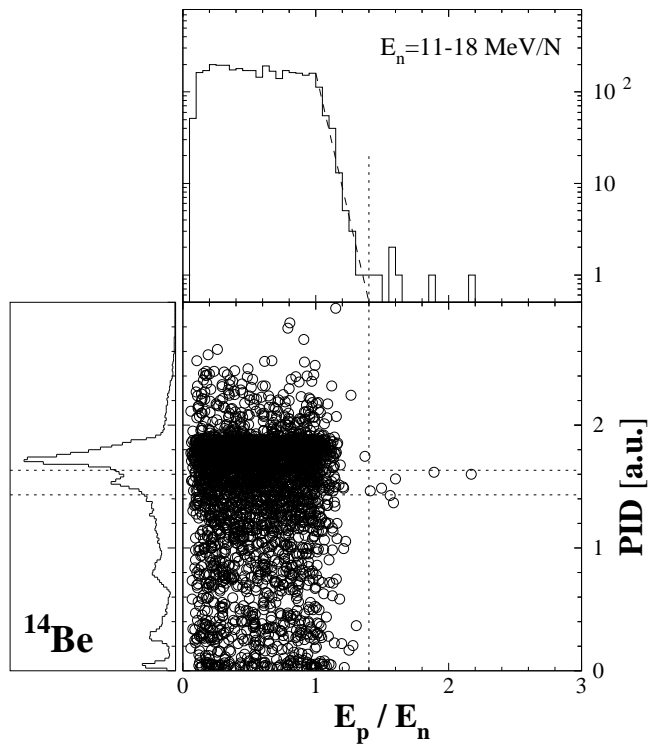,width=8.8cm}}
 \caption{Scatter plot and the projections onto both axes of the
particle identification parameter PID defined in Eq.~(\protect\ref{e:PID}) vs
$\Epn$ for the data from the reaction C($^{14}$Be,X+n). The PID projection is
displayed for all neutron energies. The dotted lines correspond to $\Epn=1.4$
and to the region centred on the $^{10}$Be peak.} \label{f:IvE}
\end{center}
\end{figure}
 
\begin{figure*}
\begin{center}
 \mbox{\psfig{file=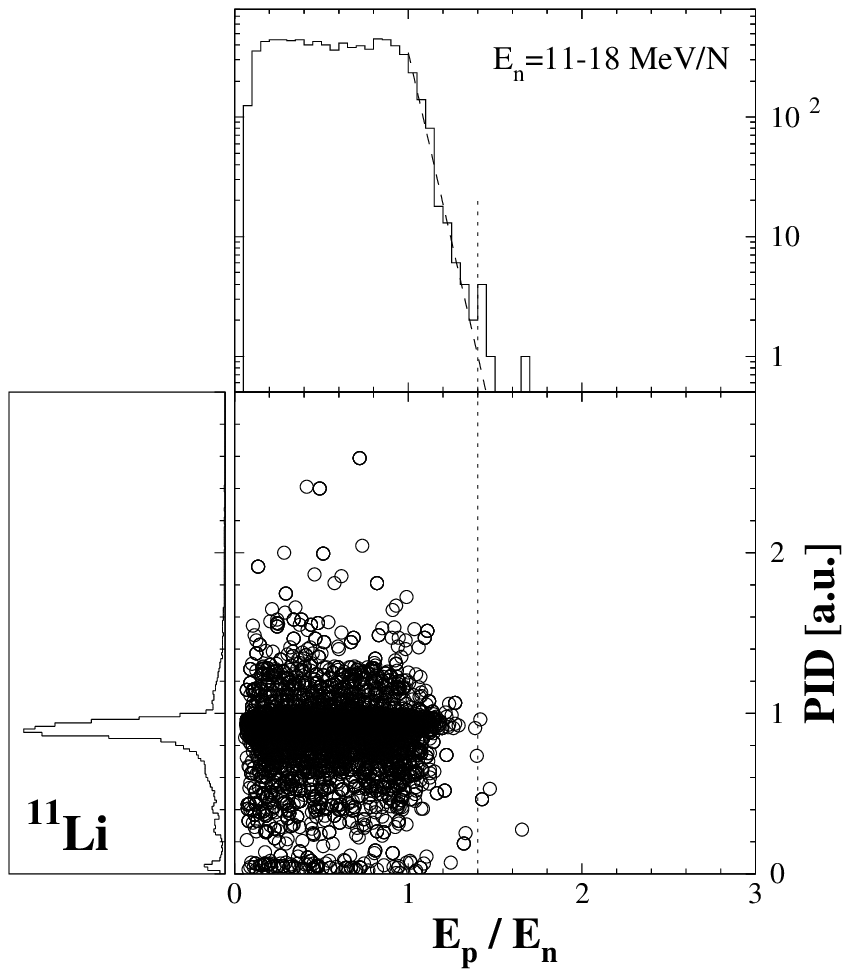,width=8.8cm}}\hspace*{5mm}
 \mbox{\psfig{file=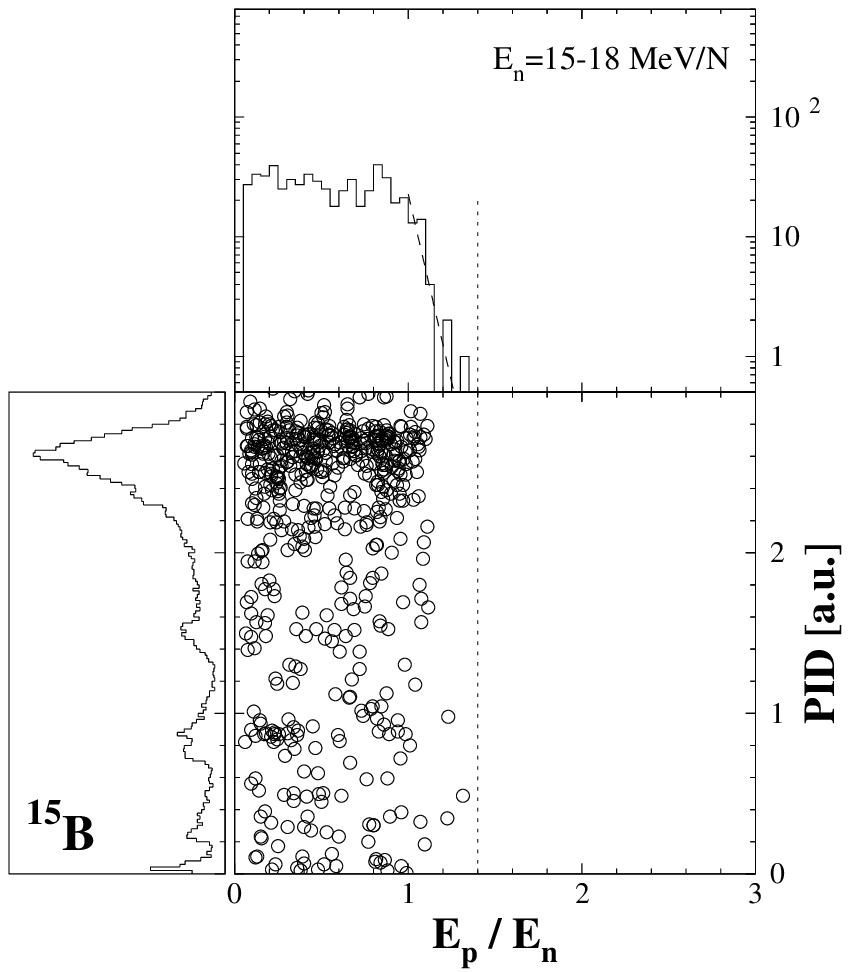,width=8.8cm}}
 \caption{Same as in Fig.~\protect\ref{f:IvE} for the data from the
reactions C($^{11}$Li,X+n) and C($^{15}$B,X+n).} \label{f:IEs}
\end{center}
\end{figure*}
 
\begin{figure}
\begin{center}
 \mbox{\psfig{file=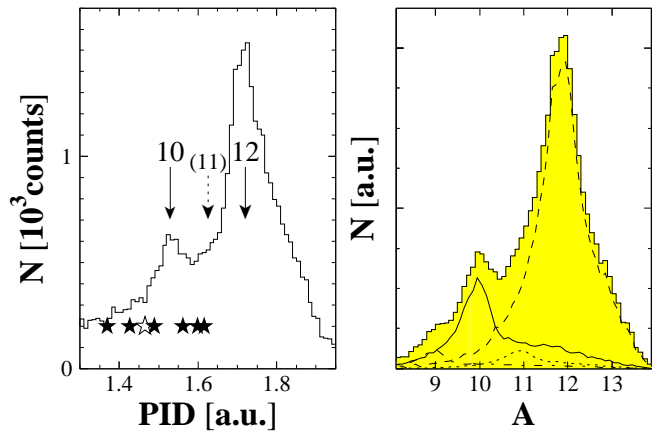,width=8.8cm}}
 \caption{Left: detail of the particle identification spectrum around
$^{10,12}$Be for the data from the reaction C($^{14}$Be,X+n); the 7 events with
$\Epn>1.4$ are denoted by the symbols. Right: results of the simulation
described in the text for the reaction ($^{14}$Be,$^{9-12}$Be); the shaded
histogram is the sum of all contributions (lines).} \label{f:PID}
\end{center}
\end{figure}

The charged fragments produced in the breakup of the beam particles were
identified using the energy loss ($\Delta{E}_{\rm{Si}}$) and residual energy
($E_{\rm{CsI}}$) signals derived from the telescope (Fig.~\ref{f:loc}). One
dimensional spectra representing the particle identification were constructed
as \cite{Lab99,Rii92}:
\begin{equation}
 {\rm{PID}} = (\Delta{E}_{\rm{Si}}+a)
              \exp\left\{-(E_{\rm{CsI}}-b)^2/2c^2\right\} \label{e:PID}
\end{equation}
The PID distribution for each beam (left panels in
Figs.~\ref{f:IvE},\ref{f:IEs}) exhibits peaks corresponding to isotopes of H,
He, Li, Be and B. The parameters $a,b,c\,$ of Eq.~(\ref{e:PID}) were adjusted
using the Be isotopes \cite{Lab99}, in which the peaks corresponding to
$^{10,12}$Be are well resolved (Figs.~\ref{f:IvE},\ref{f:PID}). The
cross-sections for the production of $^{12,11,10}$Be from the breakup of
$^{14}$Be were $460\pm40$, $85\pm15$ and $145\pm20$~mb, respectively
\cite{Lab99}.
 
The $\Epn$ distributions (upper panels in Figs.~\ref{f:IvE},\ref{f:IEs})
exhibit a general trend below 1.4: a plateau up to 1 followed by a sharp
decline, which may be fitted to an exponential distribution (dashed line). In
the region where $^A$n events may be expected to appear, $\Epn>1.4$, some 7
events ranging from 1.4 to 2.2 are observed for $^{14}$Be. However, in the case
of $^{11}$Li, despite the greater number of neutrons detected (factor of 2.4)
only 4 events appear and lie just above threshold, whilst there are no events
in the case of $^{15}$B. Turning to the coincidences with the charged fragments
produced in the breakup, the 7 events produced by the $^{14}$Be beam fall
within a region centred on $^{10}$Be. In the case of the 4 events produced in
the reactions with $^{11}$Li no similar correlation appears to exist.
 
The left panel in Fig.~\ref{f:PID} displays in more detail the region of the
particle identification spectrum for the breakup of $^{14}$Be into lighter Be
isotopes, together with the 7 events for which $\Epn>1.4$. Clearly the
resolution in PID does not allow us to unambiguously associate all of the
observed events with a $^{10}$Be fragment. However, the much higher
cross-section for this channel compared to $^{9,11}$Be does suggest that this
may be the case.
 
Reactions in the Si-CsI telescope, which was centred at zero degrees
\cite{Lab01}, also affect the PID. If $^{14}$Be traverses the target and breaks
up in the telescope, the $\Delta{E}_{\rm{Si}}$ will correspond to $^{14}$Be but
the $E_{\rm{CsI}}$ may take on a range of values depending on the interaction
point. Such events correspond to the horizontal band centred on
$\Delta{E}_{\rm{Si}}\sim2.5$ in Fig.~\ref{f:loc}. The effects of breakup into
lighter Be isotopes in the different elements of the telescope have been
simulated. The results (right panel in Fig.~\ref{f:PID}) reproduce well the
characteristics of the observed PID spectrum. Each peak exhibits tails: to the
left owing to the energy response of the CsI and to the right owing to the
energy loss of $^{14}$Be in the CsI before breakup. Thus, whilst the PID are
further broadened by the effects of the reactions in the telescope, the 7
events in question are most probably associated with a $^{10}$Be fragment.
 
As a first step towards investigating the nature of these events we have
verified that each corresponds to a well defined event in both the charged
particle and neutron detectors. As described above, in terms of the charged
fragment identification (Figs.~\ref{f:loc},\ref{f:PID}), the 7 events lie
within the region corresponding to $^{10}$Be. One of the events, however,
exhibits a very low $E_{\rm{CsI}}$, such that it is located well away from the
region of the plot where most of the yield is concentrated. This event, with
$\Epn=1.41$, is denoted by the open symbol in Figs.~\ref{f:loc},\ref{f:PID}. In
addition, it is not kinematically consistent with the breakup of $^{14}$Be into
$^{10}$Be+$^{3,4}$n, as a low energy multineutron (11--18~MeV/N) should be
associated with a high energy $^{10}$Be. In the following the discussion is
therefore limited to the 6 remaining events.
 
\begin{figure}
\begin{center}
 \mbox{\psfig{file=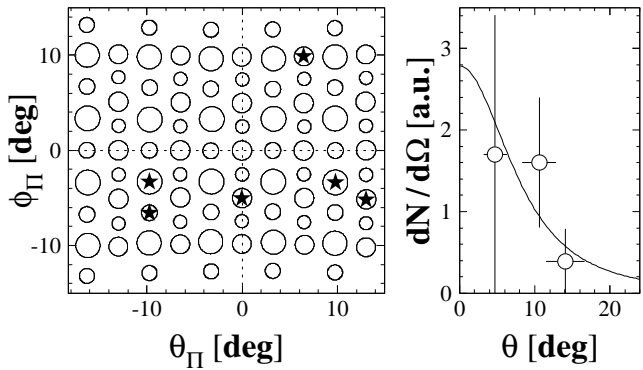,width=8.8cm}}
 \caption{Left: out-of-plane vs in-plane disposition of the DEMON
modules as viewed from the target and the location of the 6 $^{10}$Be+$^{3,4}$n
candidate events (symbols). Right: angular distribution.} \label{f:4ns}
\end{center}
\end{figure}
 
\begin{figure}
\begin{center}
 \mbox{\psfig{file=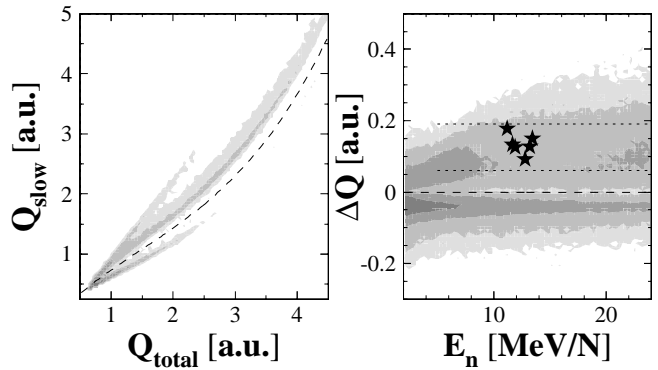,width=8.8cm}}
 \caption{Pulse-shape discrimination in DEMON for data from the reaction
C($^{14}$Be,X+n). Left: integrated slow component ($Q_{\rm{slow}}$) vs the
total integrated light output ($Q_{\rm{total}}$) for a module; the dashed line
separates the neutrons (upper group) and $\gamma$ or cosmic rays (lower group).
Right: distance in the previous plot to the dashed line as a function of the
neutron energy for all modules; the symbols correspond to the 6
$^{10}$Be+$^{3,4}$n candidate events.} \label{f:ban}
\end{center}
\end{figure}
 
Regarding the detection using DEMON, we have checked that the 6 candidate
events do not cluster in a single detector or group of detectors
(Fig.~\ref{f:4ns}). The operation of each of the detectors was then examined in
detail ---in particular the pulse-shape analysis. The standard procedure to
select the events arising from the neutrons is to compare the integrated slow
component ($Q_{\rm{slow}}$) of the light output with the total integrated
output ($Q_{\rm{total}}$) \cite{Til95}. As displayed in Fig.~\ref{f:ban},
excepting at very low $Q_{\rm{total}}$, the neutrons form a distinct group from
the $\gamma$ and cosmic-ray events. In order to present and compare the 90
modules the data have been standardised in terms of the parameter $\Delta{Q}$,
defined as the distance an event lies away from the line separating the two
groups. The data from the 90 detectors are plotted in Fig.~\ref{f:ban} and the
6 events in question are highlighted. As is clearly evident all 6 events fall
within the region corresponding to the detection of neutrons. In order to be
sure that any possible contamination by $\gamma$ and cosmic-ray events
(section~2) was minimal, the gate shown by the dotted lines in Fig.~\ref{f:ban}
was applied to the data.
 
\section{Discussion}
 
The 6 events in question thus exhibit characteristics consistent with detection
of a multineutron cluster from the breakup of $^{14}$Be. Any potential sources
of events with $\Epn>1.4$ not involving the formation of a multineutron are now
examined in detail. Each scenario must account for the fact that the 6 events
appear to be produced in association with $^{10}$Be fragments (17~\% of the
total yield) while no events appear in association with other fragments with
comparable (H-He-Li, 19~\%) or higher ($^{12}$Be, 47~\%) yields.
 
\begin{figure*}[t]
\begin{center}
 \mbox{\psfig{file=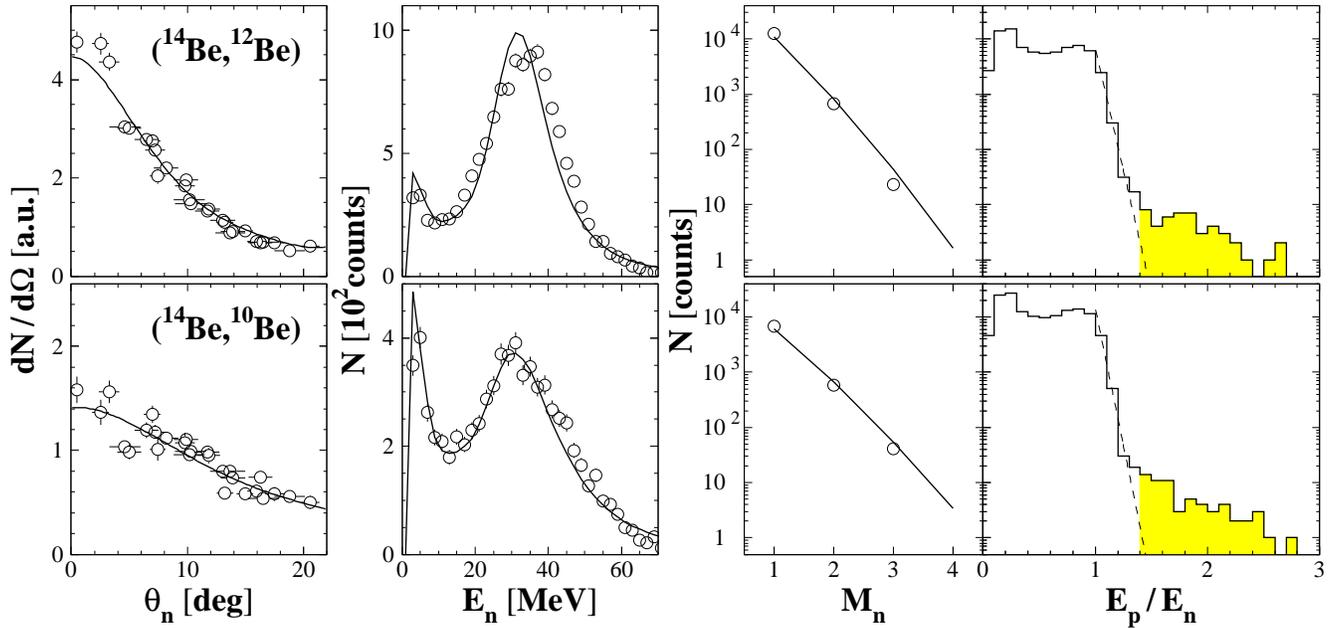,width=17.8cm}}
 \caption{Data from the reactions C($^{14}$Be,$^{12}$Be+n), upper, and
C($^{14}$Be,$^{10}$Be+n), lower panels. From left to right: distributions of
neutron angle, energy and multiplicity. The open symbols are the data, the
solid lines are results of the simulations described in the text. The histogram
in the right-most panel is for a simulation in which $10^7$ events were
generated; the dashed line is the result of a fit to an exponential for
$\Epn>1$.} \label{f:sim} 
\end{center}
\end{figure*}
 
\begin{table*}
\begin{center}
 \caption{Parameters for the simulations shown in
Fig.~\protect\ref{f:sim}: average neutron multiplicity and momentum width for
the projectile-like component; average multiplicity, energy slope and angular
width for the target-like component; and the pile-up probability deduced from
the events within the shaded area in Fig.~\protect\ref{f:sim}.} \label{t:sim}
 \begin{tabular}{lllllll} \noalign{\medskip}\hline\hline\noalign{\smallskip}
 channel & $\langle{M({\rm{Be}})}\rangle$ & $\Gamma$ [MeV/$c$] &
 $\langle{M({\rm{C}})}\rangle$ & $E_0$ [MeV] & $\theta_{1/2}$ [$^\circ$] &
 $P_{\rm{2n}}$ \\ \noalign{\smallskip}\hline\noalign{\smallskip}
 ($^{14}$Be,$^{12}$Be) & 1.4 &  85 & 1.6 & 12 & 35 & $(6\pm1)\times10^{-4}$ \\
 ($^{14}$Be,$^{10}$Be) & 2.8 & 140 & 4.0 &  8 & 40 & $(4\pm1)\times10^{-4}$ \\
 \noalign{\smallskip}\hline\hline
 \end{tabular}
\end{center}
\end{table*}
 
We first address the most probable source of events, neutron pile-up, i.e.\ the
detection for the {\em same event\/} of more than one neutron in the same
module. Alternative sources, including pile-up of a neutron and a $\gamma$-ray,
the detection in DEMON of light charged particles, and the random background
which was in principle rejected by the condition $E_{\rm{n}}>11$~MeV/N, are
also reviewed.
 
\subsection{Neutron pile-up}
 
Let us consider the possibility of pile-up, or sum events \cite{Rii92}, due to
the detection of more than one neutron in the same module. As the intrinsic
neutron detection efficiency is moderately low and the detector array highly
granular, the pile-up of just two neutrons will be the leading contribution. We
define the probability of pile-up leading to $\Epn>1.4$ for a given channel $X$
as:
$$ P_{\rm{2n}}(X) = N_{\rm{2n}}(X) \,/\, N(X) $$
where $N$ is the total number of events observed for the channel $X$+n and
$N_{\rm{2n}}$ the number with $\Epn>1.4$. A simple estimate of the number of
events expected from pile-up may be derived from channels in which no events
with $\Epn>1.4$ are observed. For example:
\begin{eqnarray*}
 P_{\rm{2n}}(\mbox{$^{12}$Be}) & < & 1/N(\mbox{$^{12}$Be}) \\
 P_{\rm{2n}}(\mbox{H-He-Li})  & < & 1/N(\mbox{H-He-Li})
\end{eqnarray*}
We assume that 2 neutrons are emitted in the $^{12}$Be channel and 4 in the
$^{10}$Be one. In the latter case, therefore, there are 3 neutrons available to
pile up. For the H-He-Li channel we assume emission of at least 4 neutrons. It
is also assumed that the pile-up distribution in $\Epn$ is flat and extends up
to $\sim3$, which means that half of these events lie above $\Epn=1.4$ (it will
be shown below that this considerably overestimates the contribution).
Therefore:
\begin{eqnarray*}
 N_{\rm{2n}}(\mbox{$^{10}$Be}) & < &
 N(\mbox{$^{10}$Be}) \times 3/2\,P_{\rm{2n}}(\mbox{$^{12}$Be}) = 0.5 \\
 N_{\rm{2n}}(\mbox{$^{10}$Be}) & < &
 N(\mbox{$^{10}$Be}) \times 1/2\,P_{\rm{2n}}(\mbox{H-He-Li}) = 0.4
\end{eqnarray*}
At most 0.5 pile-up events are expected, corresponding to a probability
$P_{\rm{2n}}(\mbox{$^{10}$Be})\sim9\times10^{-4}$.
 
In this simple estimate we do not consider neutrons from the target nucleus,
which may differ in energy, angular and multiplicity distributions for the
different reaction channels, or the non-trivial effect of the conditions
applied in the analysis, such as the limits placed on the energy $E_{\rm{n}}$.
In order to take into account these effects and how they influence the pile-up
probability in each channel, a complete Monte-Carlo simulation has been
performed employing the code MENATE \cite{fmm00,Des91}. In the $^{10,12}$Be
channels, neutrons from $^{14}$Be were described by an average multiplicity,
with a distribution ranging from 0 to 4,2 (respectively), and a Lorentzian
momentum distribution in the projectile frame with the measured width
($\Gamma$) \cite{Lab01,Lab99}. For the neutrons from the target nucleus, the
simulation also used an average multiplicity, with a distribution ranging from
0 up to 6, an energy distribution of the form $e^{-E_{\rm{n}}/E_0}$, and a
Gaussian angular distribution with a half-width $\theta_{1/2}$.
 
The different parameters were chosen so as to reproduce, in each channel, the
measured neutron energy, angular and multiplicity distributions. These
distributions (open symbols in Fig.~\ref{f:sim}) display the main differences
between the neutrons emitted in the two channels: (i) the target nucleus
contribution for $^{10}$Be is larger and exhibits a smaller slope; (ii)
neutrons in coincidence with $^{10}$Be exhibit a broader momentum distribution;
(iii) the multiplicity is higher for $^{10}$Be. The simulation (solid lines),
using the parameters listed in Table~\ref{t:sim}, provides a good overall
description of all these features.
 
The histograms in Fig.~\ref{f:sim} represent the $\Epn$ distributions obtained
after generating $10^7$ events and applying the same analysis conditions as
those applied to the data from the experiment. It should also be noted that the
simulation includes the pile-up arising from more than two neutrons interacting
in a module. The events resulting from pile-up are clearly evident for $\Epn$
above 1.4. Furthermore, as expected for pile-up, the number of events decreases
with increasing $\Epn$. Interestingly, $P_{\rm{2n}}$ for the $^{10}$Be and
$^{12}$Be channels are comparable (Table~\ref{t:sim}) despite the higher
multiplicity for the former ---this is most probably due to the enhanced
forward focussing of the halo neutrons in the latter channel. The estimates
lead to $N_{\rm{2n}}(\mbox{$^{12}$Be})=0.8$, in agreement with the experiment,
and $N_{\rm{2n}}(\mbox{$^{10}$Be})=0.2$. Assuming
$P_{\rm{2n}}\sim5\times10^{-4}$ as an average value for the other two beams,
$N_{\rm{2n}}\sim3.3$ for $^{11}$Li and $N_{\rm{2n}}\sim0.3$ for $^{15}$B, also
in agreement with experiment.
 
\begin{figure}
\begin{center}
 \mbox{\psfig{file=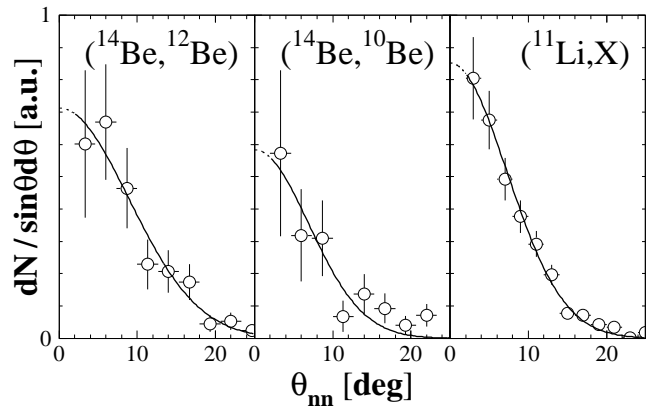,width=8.8cm}}
 \caption{Distribution of relative angle for neutron pairs from the
reactions C($^AZ$,X+2n) with $^{14}$Be and $^{11}$Li beams, selected with
conditions equivalent to those used in
Figs.~\protect\ref{f:IvE},\protect\ref{f:IEs}. The solid line is the result of
a fit to a Gaussian, which is extrapolated to $\theta_{\rm{nn}}<2^\circ$
(dotted line).} \label{f:tnn}
\end{center}
\end{figure}

Another estimate of pile-up, which does not rely on simulations, may be made.
By selecting multiplicity 2 events, the number of events in which two neutrons
interact in the same detector can be deduced from the relative-angle
distribution. Importantly, this approach includes explicitly all the effects
arising from the different conditions applied, as the experimental
distributions (Fig.~\ref{f:tnn}) have been extracted under analysis conditions
equivalent to those used to derive the results in
Figs.~\ref{f:IvE},\ref{f:IEs}. The distributions have been fitted according to
a Gaussian lineshape for $\theta_{\rm{nn}}>2^\circ$ (solid lines). The number
of pairs that may be expected to be detected in a single module can now be
estimated from the extrapolated fit (a single module on average subtended
$1^\circ$). Using the value of the fit at $1^\circ$ and assuming that half the
pairs result in $\Epn>1.4$, the estimated pile-up probability is:
$$ P_{\rm{2n}}(\mbox{$^{10}$Be})\sim12\times10^{-4} $$
 
\begin{table}
\begin{center}
 \caption{Comparison of the number of events with $\Epn>1.4$ for each
channel (1$^{\rm{st}}$ column) with the estimated number of events expected
from pile-up. The methods are based on the absence of events in the $^{12}$Be
channel (2$^{\rm{nd}}$ column), Monte-Carlo simulations (3$^{\rm{rd}}$ column),
and the relative-angle distribution of n-n pairs (4$^{\rm{th}}$ column).}
 \label{t:pup}
 \begin{tabular}{lllll} \noalign{\medskip}\hline\hline\noalign{\smallskip}
 channel & $N_{\rm{2n}}^{\rm{exp}}$ &
 $N_{\rm{2n}}^{(12)}$ & $N_{\rm{2n}}^{\rm{(sim)}}$ &
 $N_{\rm{2n}}^{\rm{(nn)}}$ \\ \noalign{\smallskip}\hline\noalign{\smallskip}
 ($^{11}$Li,X)         & 4 & $<$6.0 & $\sim$3.3 & $<$7.0 \\
 ($^{15}$B,X)          & 0 & $<$0.5 & $\sim$0.3 & $<$0.9 \\
 ($^{14}$Be,$^{12}$Be) & 0 &    --  &       0.8 & $<$1.2 \\
 ($^{14}$Be,$^{10}$Be) & 6 & $<$0.5 &       0.2 & $<$0.8 \\
 \noalign{\smallskip}\hline\hline
 \end{tabular}
\end{center}
\end{table}
 
It should be stressed that this estimate represents a very conservative upper
limit. Even if the uncertainty of the extrapolated fit at $1^\circ$ is of the
order of 20~\% for the $^{10,12}$Be channels (Fig.~\ref{f:tnn}), the number of
pile-up events with $\Epn>1.4$, as indicated by the simulations, is much less
(1/3--1/6) than half the number of double hits. This method thus suggests that
$N_{\rm{2n}}(\mbox{$^{10}$Be})<0.8$, $N_{\rm{2n}}(\mbox{$^{12}$Be})<1.2$,
$N_{\rm{2n}}<7.0$ for $^{11}$Li and $N_{\rm{2n}}<0.9$ for $^{15}$B. These
results are in line, except in the case of $^{10}$Be, with the number of
observed events. A summary of the different estimates of the number of pile-up
events is given in Table~\ref{t:pup}.
 
\subsection{Other sources of background}
 
A number of other sources of background may be postulated to account for the
excess of events with $\Epn>1.4$. Above all, any such processes must account
for the fact that these events appear to be correlated with a $^{10}$Be
fragment. One possible process that can be eliminated is the production of a
large number of $\gamma$-rays associated with the ($^{14}$Be,$^{10}$Be)
reaction channel. Pile-up of a neutron with a $\gamma$-ray could conceivably
produce events with $\Epn>1.4$, but the flight-time for such events would be
that of the $\gamma$-ray and therefore the event would be rejected.
 
Other processes involving the detection of particles not arising from reactions
of the $^{14}$Be beam particle cannot be at the origin of the events in
question, as they would not give rise to a correlation with a particular
channel. Light charged particles, for example, if present in the beam and yet
not triggering the Si beam-identification detector or being detected in the
Si-CsI telescope, should show no correlation with any particular charged
fragment. In addition, such events would be concentrated very close to the beam
axis, which is not the case for the 6 events observed here (Fig.~\ref{f:4ns}).
 
Finally, for background events in DEMON leading to $\Epn>1.4$ for an apparent
$E_{\rm{n}}<11$~MeV/N (section~2), we have considered the correlation with the
charged fragment identification (PID). The result is shown in the lower panel
of Fig.~\ref{f:Ens} for the $^{14}$Be beam data. The rate of events,
$\sim5$~\%, is independent of the PID, suggesting that such a contamination
cannot be at the origin of the observed events.
 
\subsection{Kinematical constraints and properties}
 
We have analysed the kinematics of the 6 events that are consistent with the
detection of a bound $^{3,4}$n cluster. Breakup reactions in the $^{10}$Be
channel leading to no multineutrons, a trineutron, or a tetraneutron are:
\begin{eqnarray}
 \mbox{$^{14}$Be$^*$}
 & \rightarrow & \mbox{$^{10}$Be+4n}      \label{e:5bk} \\
 & \rightarrow & \mbox{$^{10}$Be+$^3$n+n} \label{e:3bk} \\
 & \rightarrow & \mbox{$^{10}$Be+$^4$n}   \label{e:2bk}
\end{eqnarray}
with five, three and two particles in the final state, respectively. In the
average-beam-velocity frame, therefore, the relative angle ($\theta^*$) between
the neutron cluster detected and the $^{10}$Be fragment should increase with
decreasing particle number, up to a value close to $180^\circ$ for reaction
(\ref{e:2bk}).
 
\begin{figure}
\begin{center}
 \mbox{\psfig{file=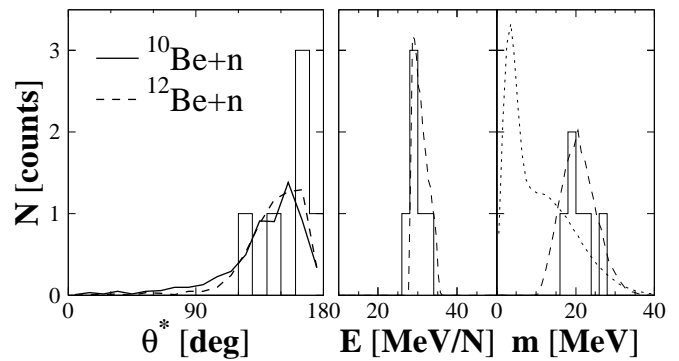,width=8.8cm}}
 \caption{Distributions for the 6 multineutron candidate events
(histograms). Left: relative angle between the Be fragment and the neutron
cluster in the average-beam-velocity frame, compared to the data for
$^{10}$Be+n (solid line) and to those for $^{12}$Be+n (dashed line); both
distributions have been normalised to 6 events. Middle-right: total kinetic
energy and invariant mass with respect to the threshold for the $^{10}$Be+$^4$n
system; the dashed lines are the result of a simulation in which the input
distribution (dotted line) has been filtered by the analysis conditions used in
Fig.~\protect\ref{f:IvE}.} \label{f:kin}
\end{center}
\end{figure}

The experimental distributions of relative angle are shown in Fig.~\ref{f:kin}:
the solid line corresponds to reaction (\ref{e:5bk}), while the dashed line is
for the $^{12}$Be channel, which is similar to reaction (\ref{e:3bk}) in that
it has three particles in the final state. Indeed, both distributions are
centred at backward angles. This is due to the selection of neutrons with
energies lower than the beam, as in the beam-velocity frame these always
correspond to neutrons emitted backwards. However, the expected trend is
observed when going from a five- to a three-particle final state, as
demonstrated by the increase in the average value from
$\langle\theta^*\rangle=139^\circ$ to $146^\circ$.
 
The relative angle between $^{10}$Be and the neutron cluster for the 6 events
is biased more closely to $180^\circ$, as expected for a back-to-back decay.
The average value is $\langle\theta^*\rangle=157^\circ$. This argues in favour
of reaction (\ref{e:2bk}), which suggests that the most probable scenario for
the events observed is the production of a tetraneutron in the breakup of
$^{14}$Be.
 
We have subsequently considered the analysis of the energies assuming that
$^{10}$Be+$^4$n breakup took place. As the tetraneutron energy has been
restricted to 11--18~MeV/N and the $^{14}$Be beam energy is around 30~MeV/N,
$^{10}$Be fragments in coincidence with $^4$n should have energies per nucleon
higher than the beam. The resulting total kinetic energy (Fig.~\ref{f:kin},
middle panel) corresponds well to the beam energy. Moreover, the reconstructed
$^{14}$Be$^*$ invariant mass might show strength at energies just above the
$^{10}$Be+$^4$n threshold ($S_{\rm4n}=5$~MeV), as was observed for $^{12}$Be+2n
\cite{Lab01}. The invariant mass distribution (Fig.~\ref{f:kin}, right panel),
however, is located far above the threshold. This is related to the low-energy
gating of the neutral fragment since, as noted above, the corresponding
$^{10}$Be fragments have high energies and thus the relative energy of the two
fragments is high. This interpretation is confirmed by the dashed lines in
Fig.~\ref{f:kin}, middle-right panels, which correspond to the results of a
simulation of reaction (\ref{e:2bk}) with a $^{14}$Be invariant mass with
strength at low energy (dotted line) and the low-energy gate on $^4$n. They are
in good agreement with the experimental distributions.
 
The events observed here may be used to provide estimates of the momentum
content, lifetime and production cross-section of the postulated tetraneutron.
A fit of the angular distribution using a Lorentzian lineshape
(Fig.~\ref{f:4ns}) leads to a momentum width of $\Gamma=90\pm60$~MeV/$c$. At
best, it can be concluded that this is compatible with the widths measured for
the single neutrons in the breakup of $^{14}$Be \cite{Lab01,Lab99}.
 
Concerning the lifetime, the events observed suggest that $^4$n is particle
stable. However, it will be unstable against $\beta$ decay because the binding
energy is limited to 3.1~MeV (section~1), which implies \cite{Aud93}:
\begin{eqnarray*}
 m(\mbox{$^4$n})-\{m(\mbox{$^3$H})+m_{\rm{n}}\} & > & \mbox{6.2~MeV} \\
 m(\mbox{$^4$n})-m(\mbox{$^4$H})  & > & \mbox{3.2~MeV}
\end{eqnarray*}
The average flight time of the 6 events from the target to DEMON is
$\sim100$~ns. This indicates that the lifetime must be of this order or longer.
 
The conditions applied in the analysis presented here make an estimate of the
total $^4$n production cross-section rather difficult. Nonetheless, if we
assume that the various gates affect the number of neutrons and tetraneutrons
in a similar manner, we can scale the cross-section measured for the production
of $^{10}$Be \cite{Lab99} by the relative yield observed and obtain
$\sigma(\mbox{$^4$n})\sim1$~mb.
 
\section{Summary}
 
A new approach to the production and detection of multineutron clusters has
been presented. The technique is based on the breakup of energetic beams of
very neutron-rich nuclei and the subsequent detection of the multineutron
cluster in liquid scintillator modules. The detection in the scintillator is
accomplished via the measurement of the energy of the recoiling proton
($E_{\rm{p}}$). This is then compared with the energy derived from the flight
time ($E_{\rm{n}}$), possible multineutron events being associated with values
of $E_{\rm{p}}>E_{\rm{n}}$.
 
The method has been applied to data from the breakup of $^{11}$Li, $^{14}$Be
and $^{15}$B by a C target. In the case of the $^{14}$Be beam, some 6 events
have been observed with characteristics consistent with the production and
detection of a multineutron cluster. Special care has been taken to estimate
the effects of pile-up. Three independent approaches were applied and it was
concluded that at most pile-up may account for some 10~\% of the observed
signal. As discussed, the most probable scenario was concluded to be the
formation of a bound tetraneutron in coincidence with $^{10}$Be.
 
The confirmation of the events observed here with a higher intensity $^{14}$Be
beam and an improved charged particle identification system, and the search for
similar events in the breakup of $^8$He, appear to be the most straightforward
steps to take in the near future. The saturation effects encountered with DEMON
at high light outputs should be reduced considerably by lowering the beam
energy and the high voltage applied to the photomultipliers. In the longer
term, searches for heavier multineutron clusters may be conducted when more
neutron-rich beams become available at intensities beyond $10^2$~pps.
 
Theoretically, multineutron clusters may provide a stringent test of our basic
understanding of the nuclear interaction. If the observed events are confirmed,
the challenge will become one of describing how neutron clusters may be bound.
Possible avenues to such an end include advances in few-body theory, changes in
parameters describing the N-N potential, or the introduction of new terms or
many-body forces that occur in proton-free environments.
 
\ack{The support provided by the staffs of LPC (in particular J.M.~Gautier,
Ph.~Desrues, J.M.~Fontbonne, L.~Hay, D.~Etasse and J.~Tillier) and GANIL
(R.~Hue, C.~Cauvin and R.~Alves~Conde) in preparing and executing the
experiments is gratefully acknowledged. This work was funded under the auspices
of the IN2P3-CNRS (France) and EPSRC (United Kingdom). Additional support from
the ALLIANCE programme (Minist\`ere des Affaires Etrang\`eres and British
Council) and the Human Capital and Mobility Programme of the European Community
(contract n$^\circ$ CHGE-CT94-0056) is also acknowledged.}

\end{document}